\documentclass{svcyclop}
\usepackage[numbers]{natbib}

\usepackage{latexsym}
\usepackage{graphicx}

\usepackage{amsmath,amssymb}

\begin{document}

\title{Wave Solutions}

\author{Ivan C.\ Christov}

\institute{Department of Mechanical and Aerospace Engineering\\
Princeton University\\
Princeton, NJ 08544, USA\\
E-mail: christov@princeton.edu}

\maketitle

\section{Overview}
In classical continuum physics, a wave is a mechanical disturbance. Whether the disturbance is stationary or traveling and whether it is caused by the motion of atoms and molecules or the vibration of a lattice structure, a wave can be understood as a specific type of solution of an appropriate mathematical equation modeling the underlying physics. Typical models consist of partial differential equations that exhibit certain general properties, e.g., hyperbolicity. This, in turn, leads to the possibility of wave solutions. Various analytical techniques (integral transforms, complex variables, reduction to ordinary differential equations, etc.) are available to find wave solutions of linear partial differential equations. Furthermore, linear hyperbolic equations with higher-order derivatives provide the mathematical underpinning of the phenomenon of dispersion, i.e., the dependence of a wave's phase speed on its wavenumber. For systems of nonlinear first-order hyperbolic equations, there also exists a general theory for finding wave solutions. In addition, nonlinear parabolic partial differential equations are sometimes said to posses wave solutions, though they lack hyperbolicity, because it may be possible to find solutions that translate in space with time. Unfortunately, an all-encompassing methodology for solution of partial differential equations with any possible combination of nonlinearities does not exist. Thus, nonlinear wave solutions must be sought on a case-by-case basis depending on the governing equation.

\section{Hyperbolic Equations}

A \emph{wave solution} is difficult to define precisely \citep{ss99}. It is undeniable, however, that we know a wave when we see it. Mathematically, \emph{partial differential equations} (PDEs) of \emph{hyperbolic} type have certain properties that allow them to possess wave solutions in a rigorous sense. For convenience, henceforth, we assume all equations are properly nondimensionalized so that we need not speak of units and dimensions. We denote by $\mathbf{x} = (x_1,\dots,x_{N_d})$ the spatial coordinates in $N_d$-dimensional Euclidean space and by $t$ the temporal variable.

\subsection{Balance Laws}

Consider the change in some field quantity $\mathbf{u} = \mathbf{u}(\mathbf{x},t)$ (e.g., displacement, velocity, electric or magnetic field, etc.) over a domain $\Omega$. When $\mathbf{u}$ is conserved, the rate of change of its volumetric average must balance the flux through (in and out of) the boundary $\partial\Omega$ of the domain and also the production/loss of $\mathbf{u}$ within the domain:
\begin{equation}
\frac{\partial}{\partial t} \frac{1}{\mathcal{V}}\int_\Omega \mathbf{u} \,dv = -  \frac{1}{\mathcal{V}}\int_{\partial\Omega} \sum_{j=1}^{N_d}\hat{n}_j\mathbf{F}_j(\mathbf{u})\,da + \frac{1}{\mathcal{V}}\int_\Omega \mathbf{R}(\mathbf{u})\,dv
\label{eq:int_blaw}
\end{equation}
where $dv$ is an infinitesimal volume element, $\mathcal{V}$ is the volume of $\Omega$, $\hat{\mathbf{n}} = (\hat{n}_1,\dots,\hat{n}_{N_d})$ is the surface normal to $\partial\Omega$ and $da$ its infinitesimal surface element. Equation~(\ref{eq:int_blaw}) is called the \emph{integral form} of the balance law for $\mathbf{u}$ with $\{\mathbf{F}_1,\dots,\mathbf{F}_{N_d}\}$ and $\mathbf{R}$, respectively, being the fluxes in the coordinate directions and the source term.

Assuming, for the sake of argument, that all mathematical prerequisites are satisfied, we can make use of the divergence theorem and require that Eq.~(\ref{eq:int_blaw}) holds for any fixed (stationary) domain $\Omega$, to obtain the \emph{differential form} of the balance law:
\begin{equation}
\frac{\partial \mathbf{u}}{\partial t} + \sum_{j=1}^{N_d}\frac{\partial}{\partial x_j}\mathbf{F}_j(\mathbf{u}) = \mathbf{R}(\mathbf{u})
\label{eq:blaw}
\end{equation}
When $\mathbf{R}\equiv\mathbf{0}$, we say that Eq.~(\ref{eq:blaw}) is a \emph{conservation law}. Furthermore, this system is \emph{hyperbolic} if the composite \emph{Jacobian} $\mathbf{J}(\mathbf{u};\hat{\mathbf{n}}) = \sum_{i=j}^{N_d}\hat{n}_j\frac{\partial\mathbf{F}_j}{\partial\mathbf{u}}(\mathbf{u})$ has only real eigenvalues and is diagonaliazable {for all} $\hat{\mathbf{n}}$ and $\mathbf{u}$ \citep[Chap.~18]{l02}. If the eigenvalues are also distinct, then the system is called \emph{strictly} hyperbolic. The eigenvalues $\lambda_l$ and right-eigenvector $\mathbf{r}_l$ may both depend on $\mathbf{u}$ as well; thus, if $\frac{\partial\lambda_l}{\partial\mathbf{u}}\cdot\mathbf{r}_l\ne0$ for all $\mathbf{u}$, the $l$th eigenpair is said to be \emph{genuinely nonlinear}, while it is \emph{linearly degenerate} when the latter condition fails to hold. It is important to note that the definition of hyperbolicity rests upon the assumption that each $\mathbf{F}_j$ does not depend on $\nabla\mathbf{u}$ (or higher-order gradients) explicitly because such a dependence would not allow us to define a Jacobian as done above. In other words, this notion of hyperbolicity only applies PDEs that can be expressed as \emph{first-order systems}.

\subsection{Linearization}

By introducing dimensionless groups of parameters, some of which are assumed to be small, one can formally develop a solution of a nonlinear equation into an asymptotic expansion. At the leading order, the equation to be solved is linear and independent of said small parameter(s). Alternatively, the problem can be ``linearized'' by evaluating all nonlinear terms at some equilibrium state, which is usually equivalent to the formal asymptotic expansion. Linearizing Eq.~(\ref{eq:blaw}) leads to
\begin{equation}
\frac{\partial \mathbf{u}}{\partial t} + \sum_{j=1}^{N_d}\mathbf{A}_j\frac{\partial \mathbf{u}}{\partial x_j} = \mathbf{B}\mathbf{u}
\label{eq:linear_blaw}
\end{equation}
where $\mathbf{A}_j := \partial\mathbf{F}_j/\partial\mathbf{u}$ and $\mathbf{B} := \partial\mathbf{R}/\partial\mathbf{u}$ represent Jacobian matrices evaluated at some (constant) equilibrium state $\mathbf{u}_0$. For convenience, we have assumed that $\mathbf{R}$ depends only on $\mathbf{u}$. Moreover, $\mathbf{u}$ has been redefined to eliminate all constant terms on both sides of Eq.~(\ref{eq:linear_blaw}).

\subsubsection{One-way Wave Equation}

In the one-dimensional case ($N_d=1$), leaving the $j=1$ subscript understood, the system in Eq.~(\ref{eq:linear_blaw}) can be decoupled whenever $\mathbf{A}$ and $\mathbf{B}$ are \emph{simultaneously} diagonalizable, i.e., $\mathbf{A} = \mathbf{S}^{-1}\boldsymbol{\Lambda}\mathbf{S}$ and $\mathbf{B} = \mathbf{S}^{-1}\boldsymbol{\Gamma}\mathbf{S}$ with $\boldsymbol{\Lambda}$ and $\boldsymbol{\Gamma}$ being diagonal matrices with the eigenvalues of $\mathbf{A}$ and $\mathbf{B}$ along their diagonals. Then, Eq.~(\ref{eq:linear_blaw}) is equivalent to
\begin{equation}
\frac{\partial \tilde u_l}{\partial t} + \lambda_l\frac{\partial \tilde u_l}{\partial x}= \gamma_l \tilde u_l
\label{eq:decoup_linear_blaw}
\end{equation}
where $\tilde{\mathbf{u}} = \mathbf{S}\mathbf{u}$, and $l$ ranges over the number of components of $\mathbf{u}$. Because Eq.~(\ref{eq:decoup_linear_blaw}) has the same form for any $l$, we drop the $l$ subscript; tildes are also left understood. 

It is now a simple exercise to show that Eq.~(\ref{eq:decoup_linear_blaw}), subject to $u(x,t=0)=\mathring{u}(x)$, has the following \emph{progressive wave solution}:
\begin{equation}
u(x,t) = e^{\gamma t}\mathring{u}(x-c t),\quad c=\lambda
\label{eq:soln_owwe}
\end{equation}
The phase velocity $c$ of the wave is determined by $\lambda$, while its attenuation is determined by $\gamma$. For $\lambda>0$ (resp.\ $\lambda<0$), $u(x,t)$ is a right (resp.\ left) traveling wave. As a result, Eq.~(\ref{eq:decoup_linear_blaw}) is often referred to as the \emph{one-way wave equation} with a source term. When $\gamma = 0$ (i.e., there is no source in the original balance law), the initial wave form propagates unchanged. Notice that when $\gamma > 0$ there is unbalanced ``production'' of $u$ in the domain, leading to the unbound growth of the solution in time. When $\gamma < 0$, on the other hand, there is unbalanced ``loss'' of $u$ within the domain, leading to $u(x,t) \to 0$ for all $x$ as $t\to \infty$. The solution given by Eq.~(\ref{eq:soln_owwe}) is illustrated in Fig.~\ref{lin_waves}(a).

An important feature of the solution given by Eq.~(\ref{eq:soln_owwe}) is its \emph{finite domains of influence and dependence}. For any given initial condition, the value $\mathring{u}(x_0)$ for all $x_0$ in the domain can travel only as far as $x=x_0+ct$ during the time interval $[0,t]$, so its domain of influence is the spatial interval $[x_0,x+ct]$, where we have assumed $c>0$ without loss of generality. Similarly, the value of any solution $u(x,t)$ for all $x$ in the domain can only be influenced by those in the spatial interval $[x-ct,x]$, which is called its domain of dependence. Both of these domains are intervals of finite length $ct$. Thus, Eq.~(\ref{eq:decoup_linear_blaw}) exhibits \emph{finite speed of propagation} of signals at a phase velocity given by $c$. Since all hyperbolic systems of first-order hyperbolic systems can be linearized and written in the form of Eq.~(\ref{eq:linear_blaw}), which can be decoupled as in Eq.~(\ref{eq:decoup_linear_blaw}), we say that hyperbolic PDEs posses wave solutions in a strict sense.

\subsubsection{Two-way Wave Equation}

Again in the one-dimensional case, suppose $\mathbf{u}$ has two components, while $\mathbf{A}$ has only off-diagonal entries. Leaving the tildes understood, Eq.~(\ref{eq:decoup_linear_blaw}) is then equivalent to the following coupled system:
\begin{subequations}\begin{align}
\frac{\partial u_1}{\partial t} + A_{12} \frac{\partial u_2}{\partial x} &= B_{11}u_1 + B_{12}u_2\\
\frac{\partial u_2}{\partial t} + A_{21} \frac{\partial u_1}{\partial x} &= B_{21}u_1 + B_{22}u_2
\end{align}\label{eq:2comp_lin_sys}\end{subequations}
Furthermore, suppose that $\mathbf{B} = \mathbf{0}$, then $u_2$ can be eliminated from the above system (by taking the $t$ derivative of the first equation and the $x$ derivative of the second equation) to obtain a scalar equation for $u \equiv u_1$:
\begin{equation}
\frac{\partial^2 u}{\partial t^2} = c^2 \frac{\partial^2 u}{\partial x^2}
\label{eq:wave}
\end{equation}
where $c := \sqrt{A_{12}A_{21}}$. Note that $A_{12}A_{21}>0$ by the assumption that the system in Eq.~(\ref{eq:2comp_lin_sys}) is hyperbolic. It can be shown that Eq.~(\ref{eq:wave}), subject to $u(x,t=0)=\mathring{u}(x)$ and $\frac{\partial u}{\partial t}(x,t=0) = \mathring{v}(x)$, has the following {progressive wave solution}:
\begin{equation}
u(x,t) = \frac{1}{2}\left[\mathring{u}(x-ct) + \mathring{u}(x+ct)\right] + \frac{1}{2c}\int_{x-ct}^{x+ct} \mathring{v}(s) \,ds
\label{eq:soln_twwe}
\end{equation}
This solution of the initial-value problem on $-\infty<x<+\infty$, $t>0$ is the celebrated \emph{d'Alembert solution} of the wave equation \citep[\S4-1]{gl88}. Note that, for any given set of initial conditions $\{\mathring{u},\mathring{v}\}$, the domain of influence at any $x_0$ is $[x_0-ct,x_0+ct]$. Meanwhile the domain of dependence of a solution at any $x$ is $[x-ct,x+ct]$. Both of these are finite intervals of length $2ct$. Since the two-way wave equation allows propagation of signals in {two} directions, the domains of influence and dependence are twice the size of those of the one-way wave equation. The solution given by Eq.~(\ref{eq:soln_twwe}) is illustrated in Fig.~\ref{lin_waves}(b).

\subsubsection{Damped Wave Equation}

\begin{figure}
\centering
\includegraphics{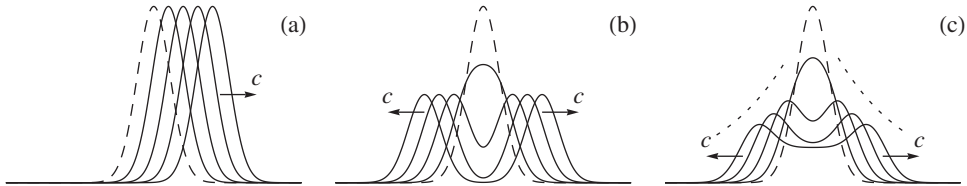}
\caption{Progressive wave solutions $u(x,t)$ (superimposed at four consecutive times), starting with a Gaussian initial condition $\mathring{u}$ (dashed curves) and $\mathring{v}=0$, of the (a) one-way wave equation ($\gamma = 0$), (b) two-way wave equation and (c) damped wave equation ($\tau=5c$, dotted curves show the exponentially-decaying amplitude of the peaks).}
\label{lin_waves}
\end{figure}

Returning to Eq.~(\ref{eq:2comp_lin_sys}), let us now take $B_{11}\ne0$ while $B_{12} = B_{21} = B_{22} = 0$. Eliminating $u_2$ between the two equations now leads to 
\begin{equation}
\frac{\partial^2 u}{\partial t^2} + \frac{1}{\tau} \frac{\partial u}{\partial t} = c^2 \frac{\partial^2 u}{\partial x^2}
\label{eq:dwe}
\end{equation}
where $c := \sqrt{A_{12}A_{21}}$ and $\tau := -1/B_{11}$. The term proportional to $\tau$, which is usually called the \emph{relaxation time}, corresponds to damping (resp.\ growth) when $\tau > 0$ (resp.\ $\tau <0$). Equation~(\ref{eq:dwe}) is also known as the \emph{telegrapher's equation}. It arises in heat conduction when the heat flux has exponentially-decaying ``memory'' of the history of the temperature gradient \citep{jp89,jp90}. Equation~(\ref{eq:dwe}) possesses a d'Alembert-type progressive wave solution \citep[\S5-5]{gl88}:
\begin{equation}
\begin{aligned}
u(x,t) = {} &\frac{1}{2}\left[\mathring{u}(x-ct) + \mathring{u}(x+ct)\right]e^{-t/(2\tau)}\\
&+ \frac{1}{4c\tau}e^{-t/(2\tau)}\int_{x-ct}^{x+ct} \mathring{u}(s)\left\{I_0\left(\frac{\rho(s) t}{2\tau}\right) + \frac{1}{\rho(s)}I_1\left(\frac{\rho(s) t}{2\tau}\right)\right\} ds\\
&+ \frac{c\tau}{2}e^{-t/(2\tau)}\int_{x-ct}^{x+ct} \mathring{v}(s)I_0\left(\frac{\rho(s) t}{2\tau}\right) ds
\end{aligned}
\label{eq:solN_dwe}
\end{equation}
where $\rho(s) = \rho(s;x,t) := \sqrt{1 - (x-s)^2/(ct)^2}$ and $I_\nu(\cdot)$ is the modified Bessel function of the first kind of order $\nu$. Clearly, the damped wave equation has the same domains of influence and dependence as the two-way wave equation. The solution given by Eq.~(\ref{eq:solN_dwe}) is illustrated in Fig.~\ref{lin_waves}(c).

\subsubsection{Standing Wave Solutions}

The two-way wave equation on a \emph{finite} domain, unlike the one-way wave equation, allows for the possibility of waves to superimpose in a way such that a \emph{standing wave} pattern forms. To illustrate this phenomenon, consider the two-dimensional two-way wave equation on a domain $\Omega$:
\begin{equation}
\frac{\partial^2 u}{\partial t^2} = c^2\mathrm{\Delta} u,\qquad \mathbf{x}\in \Omega,\quad t>0
\label{eq:wave2}
\end{equation}
where $\mathrm{\Delta} := \nabla\cdot\nabla$ is the Laplacian operator. Furthermore, due to the presence of a second-order spatial derivative, we must specify the value of $u(\mathbf{x},t)$ everywhere on the boundary $\partial\Omega$. 

For such \emph{initial-boundary-value} problems, the technique of \emph{separation of variables} \cite[Chap.~7]{gl88} can be used to seek a solution of the form $u(\mathbf{x},t) = \mathcal{X}(\mathbf{x})\mathcal{T}(t)$, where $\mathcal{X}$ and $\mathcal{T}$ are to be determined. This assumption reduces the problem of standing waves to one of finding the appropriate set of \emph{eigenfunctions} $\{\mathcal{X}_\mathbf{m}(\mathbf{x})\}$, where $\mathbf{m}=(m_1,\dots,m_{N_d})$ is an integer multi-index, that satisfy $-\mathrm{\Delta} \mathcal{X}_\mathbf{m}(\mathbf{x}) = \mu_\mathbf{m} \mathcal{X}_\mathbf{m}(\mathbf{x})$ in $\Omega$ and the appropriate boundary conditions on $\partial\Omega$. Using linearity of the PDE and the definition of eigenfunctions, the task of finding $\mathcal{T}(t)$ is reduced to solving a second-order ordinary differential equation (ODE) with constant coefficients. Then, the general solution to Eq.~(\ref{eq:wave2}) is a superposition of all eigenfunctions:
\begin{equation}
u(\mathbf{x},t) = \sum_{\mathbf{m}} \left[\alpha_\mathbf{m}\sin(\sqrt{\mu_\mathbf{m}}ct)+\beta_\mathbf{m}\cos(\sqrt{\mu_\mathbf{m}}ct)\right] \mathcal{X}_\mathbf{m}(\mathbf{x})
\label{eq:standing_wave}
\end{equation}
where the coefficients $\{\alpha_\mathbf{m},\beta_\mathbf{m}\}$ must be determined from the initial conditions $u(\mathbf{x},0)$ and $\frac{\partial u}{\partial t}(\mathbf{x},0)$ using orthogonality relations for the eigenfunctions. Unlike the progressive wave solutions above, the solution in Eq.~(\ref{eq:standing_wave}) does not depend on the wave coordinate $\mathbf{x}-ct$, rather it is \emph{standing wave}---a superposition of purely spatial patterns, given by the eigenfunctions $\{\mathcal{X}_\mathbf{m}(\mathbf{x})\}$, whose coefficients are evolving in time. 

A thorough treatise on eigenfunctions, orthogonal systems and series expansions is given by \citet[Chap.~II]{ch53}. See also the discussion of Eq.~(\ref{eq:wave2}) in \citep[Chap.~V \S5]{ch53}, where it is shown that all $\mu_\mathbf{m}\ge0$ for the Laplacian operator, justifying the form of the solution for $\mathcal{T}_\mathbf{m}(t)$ used in Eq.~(\ref{eq:standing_wave}).

\subsubsection{Dispersion}

For linear (not necessarily hyperbolic) PDEs with constant coefficients posed on an infinite domain, one can seek Fourier-mode wave solutions of the form $u(\mathbf{x},t) = e^{i\mathbf{k}\cdot\mathbf{x}-i\omega t}$. Here, we only consider the scalar case, however, the approach is readily extended to systems of linear equations \citep{cm92}. For $\mathbf{k}$ fixed, $e^{i\mathbf{k}\cdot\mathbf{x}-i\omega t}$ is called a \emph{monochromatic wave} solution. Upon substitution into the governing PDE, the problem is reduced to solving an algebraic equation of the form $\Xi(\omega,\mathbf{k}) = 0$, which is called the \emph{dispersion relation}. The solutions for $\omega$ as a function of $\mathbf{k}$, one or more of which may be complex-valued, determine the allowed wave \emph{modes} of the PDE.

As an example, consider the following one-dimensional linear PDE with constant real coefficients $a$, $b$ and $c$:
\begin{equation}
\frac{\partial u}{\partial t} + a \frac{\partial u}{\partial x} + b \frac{\partial^2 u}{\partial x^2} + c\frac{\partial^3 u}{\partial x^3} = 0
\label{eq:linear_disp}
\end{equation}
The resulting dispersion relation is
\begin{equation}
\Xi(\omega,k) = -i\omega + aik + b(ik)^2 + c(ik)^3 = 0
\end{equation}
Solving for $\omega$, we immediately obtain $\omega(k) = a k  + i bk^2 - ck^3$, which can be interpreted as follows. For a solution of the form $e^{i\mathbf{k}\cdot\mathbf{x}-i\omega t}$, the \emph{phase velocity} (sometimes denoted as $c_\mathrm{p}$) has the natural definition $c := \Re[\omega]\mathbf{k}/|\mathbf{k}|^2 = \Re[\omega]/k$ for one-dimensional propagation. Clearly, the third-order spatial derivative in Eq.~(\ref{eq:linear_disp}) leads to a $c_\mathrm{p}$ that is not independent of $k$, i.e., modes with different wave numbers have different frequencies (and, as a result, different phase velocities). This is the phenomenon of \emph{dispersion}. Meanwhile, the second-order spatial derivative  in Eq.~(\ref{eq:linear_disp}) leads to $\omega$ having an imaginary part, which results in either  growth ($b>0$) or decay ($b<0$) of the mode because $e^{-i\omega t} = e^{\Im[\omega]t}e^{-i\Re[\omega]t}$. If $\Re[\omega]=0$ (e.g., $a=c=0$ in Eq.~(\ref{eq:linear_disp})), then the PDE does \emph{not} possess wave solutions, strictly speaking.

To fully describe dispersive phenomena, the concept of \emph{group velocity} $c_\mathrm{g} := \nabla_\mathbf{k}\Re[\omega(\mathbf{k})]$ must also be introduced. For one-dimensional propagation, we simply have $c_\mathrm{g} = d\Re[\omega]/dk$. To understand the difference between phase and group velocities, consider a superposition of two one-dimensional monochromatic waves with nearby wave numbers given by $k$ and $k+\Delta k$, where $\Delta k \ll 1$. Then, in a dispersive medium, we have
\begin{equation}
u(x,t) = e^{i[kx-\omega(k) t]} + e^{i[(k+\Delta k)x-\omega(k+\Delta k) t]} \approx e^{ik\{x-[\omega(k)/k] t\}} \left(1 + e^{i\Delta k\{x-[d\omega/dk]t\}} \right)
\label{eq:group_vel}
\end{equation}
where we have used the truncated Taylor expansion $\omega(k+\Delta k) \approx w(k) + \frac{d\omega}{dk}\Delta k$ and taken $\omega(k)$ to be real without loss of generality. The first term on the right-hand side of Eq.~(\ref{eq:group_vel}) represents a wave traveling with phase velocity $c_\mathrm{p} = \omega(k)/k$. However, due to dispersion (i.e., the fact that $\omega(k)/k$ is not constant or, equivalently, $d\omega/dk\ne0$), the wave now has a modulated envelope that travels with its own phase velocity, namely, the group velocity $c_\mathrm{g} = \frac{d\omega}{dk}$. Thus, we arrive at the following intuitive definitions: the phase velocity describes the coherent propagation of a \emph{single} mode with wavenumber $k$, while the group velocity describes the coherent propagation of a \emph{group} of modes with nearby wavenumbers in the range $[k-\Delta k, k+\Delta k]$.

In general, if we were to project the initial condition onto the Fourier modes, use the dispersion relation to eliminate $\omega$ in favor of $\mathbf{k}$ and ``sum'' over the continuum of wave numbers (i.e., integrate over $\mathbf{k}$), then we would obtain the general solution to the linear PDE. What we have just described is the Fourier transform as a solution technique on an unbounded domain. A thorough discussion of linear dispersive wave phenomena is given by \citet[Chap.~11]{w74}.

An important application of dispersion relations is in the study of \emph{instability}. When a nonlinear PDE is linearized the coefficients carry some meaning about an equilibrium state (as discussed above). If the dispersion relation allows for solutions for $\omega$ with $\Im[\omega]>0$ for certain combinations of parameters (e.g., $b>0$ for Eq.~(\ref{eq:linear_disp})), then exponential growth of a Fourier mode can occur. Seeking to determine whether such modes exist is the subject of \emph{linear stability analysis}, which has important implications in the theory of thermal stresses \citep[Chap.~9]{nht03}. In practice, such an analysis must also take into account specific boundary conditions on finite domains.

\subsection{Nonlinear Solution Types}

For simplicity of presentation, let us restrict to one-dimensional scalar conservation laws:
\begin{equation}
\frac{\partial u}{\partial t} + \frac{\partial}{\partial x}F(u) = 0
\label{eq:1d_claw}
\end{equation}
All equations of this form, subject to some given initial data, can be solved (at least formally) by the method of characteristics \citep{ch62,l06}.

\subsubsection{Method of Characteristics}

Let us first rewrite Eq.~(\ref{eq:1d_claw}) in \emph{quasilinear} form:
\begin{equation}
\left\{\frac{\partial}{\partial t} + F'(u)\frac{\partial}{\partial x}\right\} u = 0
\label{eq:1d_claw_ql}
\end{equation}
The physical interpretation of Eq.~(\ref{eq:1d_claw_ql}) is that $u$ is advected by the ``velocity field'' $F'(u)$ and, since this ``material derivative'' vanishes, $u$ does not change with time in the co-moving frame. Therefore, for a given initial condition $u(x_0,0) = \mathring{u}(x_0)$, $u(x,t) = \mathring{u}(x_0)$ for all $x$ and $t$ such that $x=X(t)$, where $X(t)$ is the solution of the initial-value problem
\begin{equation}
\frac{dX}{dt} = F'(u(X(t),t)),\quad X(0) = x_0
\label{eq:characs}
\end{equation}
The curves $X(t)$ are a called the \emph{characteristics} of Eq.~(\ref{eq:1d_claw}). The analysis is easily extended to systems of hyperbolic conservation laws by using the fact that the Jacobian $\partial\mathbf{F}/\partial\mathbf{u}$ is assumed diagonalizable, thus the system of equations can be decoupled locally by a change of dependent variables, as we showed earlier for linear PDEs.

Since $u$ is conserved along the characteristic curves, we easily see that the general solution of Eq.~(\ref{eq:characs}) is
\begin{equation}
X(t) = x_0 + F'(\mathring{u}(x_0))t
\label{eq:charac}
\end{equation}
To evaluate $u(x,t)$ we have to solve the (possibly) nonlinear equation $x = X(t)$ for $x_0$; i.e., we have to \emph{trace back} along the characteristic. Notice that the characteristics given by Eq.~(\ref{eq:charac}) are, in fact, \emph{lines} because $F'$ does not explicitly depend on $x$ or $t$.

\begin{figure}
\centering
\includegraphics{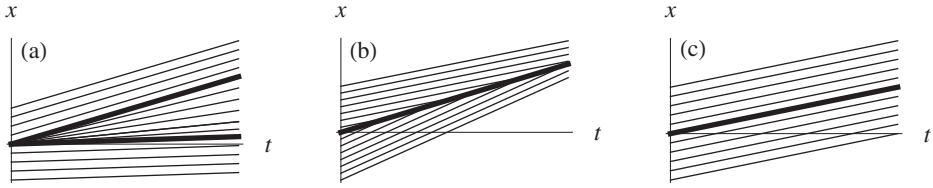}
\caption{Characteristics in the $(x,t)$ plane corresponding to a Riemann problem initial condition for (a) a rarefaction wave solution (leading and trailing edges of the rarefaction fan are bold), (b) a shock wave solution (crossing characteristics are replaced by bold shock position) and (c) a contact discontinuity wave solution (separating characteristic is bold).}
\label{claw_waves}
\end{figure}

Depending on the functional form of $F'$, different characteristic curves can have different slopes.  For simplicity, consider the case of $\mathring{u}$ being a step function. This initial-value problem for Eq.~(\ref{eq:1d_claw}) is termed the \emph{Riemann problem}. Then, depending on the left and right values $\mathring{u}^\mp := \lim_{x\to0^\mp} \mathring{u}(x)$ of this step-function initial condition, three distinct types of characteristics exist as shown in Fig.~\ref{claw_waves} \citep[see also][Chap.~2]{w74}. 

First, in Fig.~\ref{claw_waves}(a), if $F'(\mathring{u}^+)>F'(\mathring{u}^-)$, characteristics with slopes that continuously vary between $F'(\mathring{u}^+)$ and $F'(\mathring{u}^-)$ must emerge, which gives rise to a \emph{rarefaction fan}. It can be shown that $u(x,t)$ must vary continuously between $\mathring{u}^+$ and $\mathring{u}^-$ in the fan. Second, in Fig.~\ref{claw_waves}(b), if $F'(\mathring{u}^+)<F'(\mathring{u}^-)$, then the characteristics carrying these two values cross. Beyond such crossings, the solution $u(x,t)$ is multivalued. To resolve this difficulty, a \emph{shock wave} is introduced, i.e., a solution that transitions discontinuously between the values $\mathring{u}^\mp$ where the characteristics cross. Third, in Fig.~\ref{claw_waves}(c), if $F'(\mathring{u}^+)=F'(\mathring{u}^-)$ but $\mathring{u}^+ \ne \mathring{u}^-$, then the characteristics remain parallel but the one emanating from $x_0=0$ separates the two (initial) values of $u$ for all time. Such a wave is called a \emph{contact discontinuity} and is encountered in hyperbolic systems of conservation laws for which an eigenpair is linearly degenerate.

\subsubsection{Rarefaction Waves}

Since all initial condition emerging from a step-function have $x_0=0$, then we deduce from Eq.~(\ref{eq:charac}) that the characteristics in a rarefaction fan satisfy $x = F'(u(x,t))t$, or $F'(u(x,t))=x/t$. Since $F'$ is a function of $u$ alone, it follows that $u$ must be a function of the similarity variable $x/t$. In other words, rarefaction waves are self-similar solutions of Eq.~(\ref{eq:1d_claw}) in terms of $x/t$.

\subsubsection{Shock Waves}

When the characteristics cross each other, the solution jumps discontinuously between two values. Hence, it does not posses finite derivatives across the shock, and cannot satisfy Eq.~(\ref{eq:1d_claw}) as written. Thus, more generally, the concept of a \emph{weak solution} $\mathbf{u}(\mathbf{x},t)$ of Eq.~(\ref{eq:blaw}) can be introduced. Weak solutions satisfy
\begin{equation}
\int_0^T\int_\Omega \left\{ \frac{\partial \mathbf{u}}{\partial t} + \sum_{j=1}^{N_d}\frac{\partial}{\partial x_j}\mathbf{F}_j(\mathbf{u}) - \mathbf{R}(\mathbf{u}) \right\} \varphi(\mathbf{x},t) \,dv\,dt = \mathbf{0}
\label{eq:weak}
\end{equation}
for all ``smooth enough'' $\varphi(\mathbf{x},t)$ that vanish identically outside the finite space-time domain $\Omega\times(0,T)$. All strong solutions, i.e., those that satisfy Eq.~(\ref{eq:blaw}) directly, are also weak solutions. However, weak solutions are not unique as there can be many that satisfy Eq.~(\ref{eq:weak}) and the given initial/boundary conditions.

To determine the \emph{unique} weak solution, one appeals to the \emph{Lax condition} \citep{fl71}. For a system such as Eq.~(\ref{eq:1d_claw}), this condition states \citep[Chap.~10 \S3]{l06} that a shock is physical if there exists an $l$ such that
\begin{equation}
\lambda_l(\mathbf{u}^-) > V > \lambda_l(\mathbf{u}^+)
\label{eq:Lax_cond}
\end{equation}
while 
\begin{equation}
\lambda_{l-1}(\mathbf{u}^-) < V < \lambda_{l+1}(\mathbf{u}^+)
\end{equation}
In other words, the Lax condition states that a shock is physical only if ``information'' travels along the characteristics \emph{into} (and not out of) it. When this condition is satisfied, a shock (discontinuity) propagates with velocity $V$ in the $l$th equation of the system. For $l=1$, only Eq.~\eqref{eq:Lax_cond} is needed and $\lambda_l(\cdot)\equiv F'(\cdot)$.

For example, the violation of the Lax condition can be used to show that temperature shock waves (under the Maxwell--Cattaneo theory) \emph{cannot} propagate into a rigid heat conductor with temperature-dependent conductivity when the temperature dependence acts to \emph{decrease} the conductivity with respect to a reference value \citep{rbmb08}.

It remains to determine the shock velocity $V$. This is accomplished by returning to the integral form of the balance law, i.e., Eq.~(\ref{eq:int_blaw}) and requiring it holds true for a shock connecting two {constant} states $\mathbf{u}^+$ and $\mathbf{u}^-$. The result is the \emph{Rankine--Hugoniot ``jump'' conditions}:
\begin{equation}
V(\mathbf{u}^- - \mathbf{u}^+) = \mathbf{F}(\mathbf{u}^-) - \mathbf{F}(\mathbf{u}^+)
\label{eq:RH}
\end{equation}
The latter set of equations can, in theory, be solved for $V$. We expect that there exist as many (not necessarily distinct) solutions for $V$ as there are components of $\mathbf{u}$.
Now, if $\mathbf{F}(\mathbf{u})$ admits the decomposition $\mathbf{F}(\mathbf{u}) = \mathbf{A}\mathbf{u}$ with $\mathbf{A}$ \emph{independent} of $\mathbf{u}$ (as is the case if the nonlinear system is linearized), then Eq.~(\ref{eq:RH}) reduces to
\begin{equation}
V(\mathbf{u}^- - \mathbf{u}^+) =  \mathbf{A}(\mathbf{u}^- - \mathbf{u}^+)
\end{equation}
whence the velocity of a discontinuity in the $l$th component of $\mathbf{u}$ is the $l$th eigenvalue of $\mathbf{A}$. For weak shocks of the scalar conservation law in Eq.~(\ref{eq:1d_claw}), i.e., $|u^- - u^+|\ll1$, an approximation can be made: $V = \tfrac{1}{2}[F'(u^-) + F'(u^+)] + \mathcal{O}(|u^- - u^+|^2)$ \citep[\S2.6]{w74}. Furthermore, note that the source term $\mathbf{R}(\mathbf{u})$ does not enter the Rankine--Hugoniot conditions.

For example, \citet{cj10} used the Rankine--Hugoniot conditions to find the velocity of a \emph{heat pulse} (i.e., a shock wave in the temperature) propagating in a rigid stationary heat conductor with weakly temperature-dependent conductivity under the Maxwell--Cattaneo theory. They showed that 
\begin{equation}
V/c_0 = \sqrt{1 + \frac{\epsilon}{2} \theta^-}
\label{eq:heat_shock}
\end{equation}
where $\theta$ denotes a relative temperature, it has been assumed that $\theta^+=0$, $c_0$ is the speed of infinitesimal temperature waves, and $\epsilon$ is a dimensionless parameter that determines the ``strength'' of the temperature dependence. Equation~(\ref{eq:heat_shock}) illustrates an important fact about nonlinear shock wave solutions: the speed depends on the amplitude. This dependence is often inextricable, leading to the impossibility of deriving \emph{closed-form explicit} expressions for both the location and magnitude of a shock wave.

\section{Parabolic Equations}

Another class of PDEs relevant to problems in heat conduction and thermal stresses is nonlinear reaction-diffusion equations of the form
\begin{equation}
\frac{\partial \theta}{\partial t} = \nabla\cdot\left[D(\theta)\nabla \theta\right] + R(\theta)
\label{eq:react_diff}
\end{equation}
where $\theta(\mathbf{x},t)$ is the temperature change relative to some reference value, $D(\theta)$ is the temperature-dependent diffusivity of heat, and $R(\theta)$ represents either sources/sinks of heat in the domain or production/loss of heat due to a chemical reaction. For certain functional forms of $D(\theta)$ and $R(\theta)$, Eq.~(\ref{eq:react_diff}) may posses solutions that behave like waves.

\subsection{Solution Types}

A standard technique for finding wave solutions of parabolic equations is called seeking a \emph{traveling wave solution} (TWS). The procedure is as follows: in the one-dimensional context with $\theta$ as the dependent variable, set $\theta(x,t) = \Theta(x-ct)$ and determine what $\Theta$ and $c$ must be. The assumption that $\theta$ is a function of only the combined coordinate $x-ct$ reduces a PDE like Eq.~(\ref{eq:react_diff}) to an ODE. As boundary conditions one imposes the growth or the value(s) of $\Theta$ as $|x|\to\infty$. In multiple dimensions, the TWS ansatz is $\theta(\mathbf{x},t)=\Theta(\mathbf{k}\cdot\mathbf{x}-ct)$, where $\mathbf{k}$ is the \emph{a priori} unknown direction of propagation of the wave.

Though a TWS is a function of the \emph{wave coordinate} $x-ct$, the governing equation lacks hyperbolicity (and, hence, finite domains of influence and dependence), so it may be more appropriate to call such phenomena traveling \emph{patterns} rather than waves. In other words, parabolic PDEs may possess \emph{specific} traveling patterns, however, it is not always possible to find solutions in the form of traveling disturbances for \emph{general} classes of initial conditions. Yet, some (degenerate) parabolic equations admit stable \emph{compact} TWSs, i.e., TWSs that vanish outside of a finite spatial interval; such a TWS necessarily has finite domains of influence and dependence. \citet{gk04} provide an advanced mathematical treatment of TWSs of nonlinear parabolic PDEs. A large catalog of solution types, including a variety of other solution techniques, is documented by \cite{pz03}.

\subsubsection{Temperature-dependent Conductivity}

Consider heat transfer in a rigid, stationary one-dimensional heat conductor with weak temperature-dependent conductivity. This leads to $D(\theta) = 1 + \epsilon \theta$ \citep{cj59}, where $|\epsilon|\ll1$. Then, in the absence of production/loss of heat (i.e., $R(\theta)=0$), Eq.~(\ref{eq:react_diff}) becomes
\begin{equation}
\frac{\partial \theta}{\partial t} = \frac{\partial }{\partial x}\left[(1+\epsilon \theta)\frac{\partial \theta}{\partial x}\right]
\end{equation}
The TWS of the latter PDE for $\epsilon<0$ is given by \citet{cj10}:
\begin{equation}
\theta(x,t) = \frac{1}{\epsilon}\begin{cases}
-1, & \xi\leq \xi^*\\
W_0\left(\epsilon e^{\epsilon-c\xi}\right), & \xi > \xi^*
\end{cases}
\label{eq:tws_fourier}
\end{equation}
where $\xi = x-ct$, $W_0(\cdot)$ denotes the principal branch of the \emph{Lambert $W$-function} \citep{cetal96}, $\xi^* = c^{-1}(1 + \epsilon + \ln|\epsilon|)$, and any $c>0$ is allowed. Since the domain on which $\theta$ is non-constant is semi-infinite, the TWS given by Eq.~(\ref{eq:tws_fourier}) represents a \emph{semicompact} wave \citep{dgs07} with a wavefront $x = \xi^* + ct$ propagating to the right with phase velocity $c$. The solution given by Eq.~(\ref{eq:tws_fourier}) is illustrated in Fig.~\ref{nlin_waves}(a).

\subsubsection{Nonlinear Heat Equation}

At temperatures near absolute zero, the departure from a constant conductivity can be severe. It has been proposed that, as a result, the diffusivity $D(\theta) = \theta^p$ for metals at such low temperatures \citep{wc75}. Then, for one-dimensional heat transfer in a rigid, stationary conductor in the absence of heat production/loss:
\begin{equation}
\frac{\partial \theta}{\partial t} = \frac{\partial }{\partial x}\left(\theta^p\frac{\partial \theta}{\partial x}\right)
\label{eq:nl_heat}
\end{equation}
This PDE is sometimes referred to as the \emph{porous medium equation} because it arises in the study of filtration of a Newtonian fluid through a porous medium; it has a well-developed mathematical theory \citep{v07}.

Equation~(\ref{eq:nl_heat}) admits the TWS \citep[Chap.~X, \S3]{zr67}:
\begin{equation}
\theta(x,t) = \begin{cases}
\left[-pc(x-ct)\right]^{1/p}, & x<ct\\
0, & x \ge ct
\end{cases}
\label{eq:nl_heat_soln}
\end{equation}
for any real $p,c>0$. The TWS given by Eq.~(\ref{eq:nl_heat_soln}) is once again a semi-compact wave but notice the functional form is distinctly different from that in Eq.~(\ref{eq:tws_fourier}), showing the case-by-case nature of wave solutions to nonlinear parabolic PDEs. If a constraint on the heat flux $q=\int_{-\infty}^{-\infty} \theta(x,t)\,dx$ is imposed, then the front location is not $x=ct$ (for any $c>0$) but $x\propto t^{\varsigma(p)}$ \citep{zr67}. The solution given by Eq.~(\ref{eq:nl_heat_soln}) is illustrated in Fig.~\ref{nlin_waves}(b).

\subsubsection{Fisher-KPP Equation}

\begin{figure}
\centering
\includegraphics{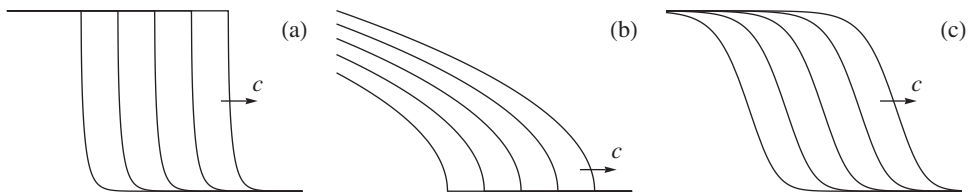}
\caption{Traveling wave solutions $\theta(x,t)$ (superimposed at five consecutive times) of (a) the temperature-dependent-conductivity heat equation, (b) the nonlinear heat equation ($p=2$) and (c) the Fisher-KPP equation.}
\label{nlin_waves}
\end{figure}

Now, consider the case in which the diffusivity is constant, i.e., $D(\theta) = 1$, but introduce a saturating reaction term with logistic-type nonlinearity: $R(\theta) = \theta(1-\theta)$. Equation~(\ref{eq:react_diff}) now becomes the so-called Fisher--Kolmogorov--Petrovskii--Piskunov (Fisher-KPP) equation:
\begin{equation}
\frac{\partial \theta}{\partial t} = \frac{\partial^2\theta}{\partial x^2} + \theta(1-\theta)
\label{eq:fkpp}
\end{equation}
which has the following TWS \citep{az79}: 
\begin{equation}
\theta(x,t) = \frac{1}{[1 + e^{(x-x_0-ct)/\sqrt{6}}]^2},\quad c = \frac{5}{\sqrt{6}}
\label{eq:FKPP_soln}
\end{equation}
where $x_0$ is an arbitrary constant used to set the initial front position. Unlike the solutions in Eqs.~(\ref{eq:tws_fourier}) and (\ref{eq:nl_heat_soln}), the TWS given by Eq.~(\ref{eq:FKPP_soln}) is \emph{not} semicompact. Rather, it is a smooth wavefront, connecting two unequal constant states at $x\to\pm\infty$. Also unlike the previous two examples, only one phase velocity $c=5/\sqrt{6}$ is allowed under Eq.~(\ref{eq:fkpp}). The solution given by  Eq.~(\ref{eq:FKPP_soln}) is illustrated in Fig.~\ref{nlin_waves}(c).

\subsection{Further Reading}

Classical references, focusing on (amongst other things) the theory of wave solutions to problems in classical physics (e.g., gas dynamics, water waves, etc.), include \citet{cf48}, \citet{mf53}, \citet{ch53,ch62} and \citet{w74}. An introductory textbook by \citet{b88} is available on the subject, while \citet{cm92} cover wave propagation inhomogeneous media in detail. A review of the wave theory of heat conduction was written by \citet{jp89,jp90}. The topic has been reexamined by \cite{s11} to include recent results. Wave solutions in the study of thermoelasticity with finite wave speeds are discussed by \citet{ios10}. 

Modern textbooks on hyperbolic PDEs and wave solutions are necessarily focused on the numerical as well as the theoretical aspects. \citet{l02} and \citet{t07} masterfully blend these topics. The set of lecture notes by \citet{l06}, who developed much of the early theory, provides a gentle introduction to the fundamental mathematical topics related to hyperbolic wave phenomena. 

A class of wave solutions to hyperbolic equations not discussed above is those that are smooth in the field variables but may suffer a jump discontinuity in a higher derivative. Known as \emph{acceleration waves} in fluid mechanics and \emph{temperature-rate waves} in heat conduction, these ``weak discontinuities'' are discussed in some detail by \citet[Chap.~6 \S9]{b88}, \citet[Chap.~10]{ios10} and \citet[Chap.~4]{s11}.

\emph{Solitons} are another important category of wave solutions that were not discussed above. They arise from a balance of nonlinearity and {dispersion}. In general, solitons are unlike the traveling wave solutions of the nonlinear parabolic PDEs discussed above because (a) a soliton's wave speed depends on its amplitude and (b) two solitons can pass through each other (``interact'') remaining unchanged. It has been suggested that soliton-like temperature waves may exist in nonlinear rigid conductors at low temperatures under the theory of \citet{hi95}. A detailed introduction to the mathematical theory of solitons is presented by \citet{as81}, while \citet{dp06} give an overview of their physics and applications.

\section{Cross-references}
Ordinary differential equations, Partial differential equations, Laplace and Fourier transforms, Special functions, Hyperbolic Heat Conduction Equation, Thermoelastic waves, Propagation of wavefronts in thermoelastic media, Piezothermoelasticity with finite wave speeds, Wave propagation in Coupled and Generalized Thermoelasticity



\end{document}